\documentclass[%
 reprint,
 amsmath,amssymb,
 aps,
]{revtex4-1}

\usepackage{amsthm}
\usepackage{graphicx}
\usepackage{dcolumn}
\usepackage{bm}
\usepackage{float}

\begin{document}

\preprint{APS/123-QED}

\title{Semi-device-independent randomness expansion with  partially free random sources}

\author{Yu-Qian Zhou$^{1}$}
 \author{Hong-Wei Li$^{2}$ }
\author{Yu-Kun Wang$^{1}$ }%
\author{Dan-Dan Li$^{1}$ }
\author{Fei Gao$^{1}$}
 \email{gaofei\_bupt@hotmail.com}
 \author{Qiao-Yan Wen$^{1}$}

\affiliation{%
 $^{1}$State Key Laboratory of Networking and Switching Technology, Beijing University of Posts and Telecommunications, Beijing, 100876, China\\
$^{2}$Key Laboratory of Quantum Information, University of Science and Technology of China, Hefei, 230026, China}

\date{\today}

\begin{abstract}
By proposing device-independent protocols, S. Pironio  \emph{et al.} [Nature \textbf{464}, 1021-1024 (2010)] and R. Colbeck \emph{et al.} [Nature Physics \textbf{8}, 450-453 (2012)] proved that new randomness can be generated by using  perfectly free random sources or partially free ones as seed. Subsequently, Li \emph{et al.} [Phys. Rev. A \textbf{84}, 034301 (2011)] studied this topic in the framework of semi-device-independent  and proved that new randomness can be obtained from perfectly free random sources. Here we discuss whether and how partially free random sources bring us new randomness in semi-device-independent scenario.
We propose a semi-device-independent randomness expansion protocol with partially free random sources, and obtain the condition that the partially free random sources should satisfy to generate new randomness. In the process of analysis, we acquire a new 2-dimensional quantum witness. Furthermore, we get the analytic relationship between the generated randomness and the 2-dimensional quantum witness violation.


\begin{description}
\item[PACS numbers] 03.67.Ac, 05.40.-a

\end{description}
\end{abstract}

\pacs{Valid PACS appear here}
\maketitle

\section{\label{sec:level1}Introduction\protect}

Perfectly free random bits have both theoretical and practical significance. In the aspect of theory, perfectly free random bits are beneficial for the foundation of physical theory to establish symmetries [1]. In practical applications, perfectly free random bits could be used in many important fields, especially in cryptography. Almost all the security of cryptographic protocols depends on perfectly free random bits. For example, in the well known BB84 protocol [2], the security will be seriously limited once an eavesdropper uses partially free random bits to replace the perfectly free ones [3].

Recently, the studies of device-independent (DI) and semi-device-independent (SDI) protocols have attracted a lot of attention. Here, DI means that no assumption is made on the devices used to perform protocols [4]. Subsequently,  M. Paw{\l}owski introduced the concept of SDI meaning that the devices in protocols are noncharacterized except the tight bound of the dimension of the potential required systems [5].

Randomness expansion is the protocol in which random sources are used as seed to produce new randomness. Recently, R. Colbeck proposed a DI randomness expansion protocol based on the tripartite GHZ-type entangled states [6] and S. Pironio \emph{et al.} proposed the protocol based on Bell inequality violation [7]. These results  demonstrated that perfectly free sources can be expanded in the framework of DI.
In 2012, R. Colbeck \emph{et al.} showed that new randomness can also be obtained by using partially free bits as seed in the framework of DI (more precisely, the partially free bits can be amplified to make perfectly free ones and this process also is a DI randomness amplification protocol) [1].
Subsequently, Li  \emph{et al.} studied this interesting topic in the framework of SDI and proved that new randomness can be produced from perfectly free sources by presenting SDI randomness expansion protocols [8, 9]. Therefore, whether and how partially free sources bring us new randomness in the framework of SDI is a problem about which people may be curious.


Here, we demonstrate that new randomness can be generated from partially free sources in the SDI scenario by proposing a SDI randomness expansion protocol with partially free sources. Different from the assumption that $\varepsilon_{1}=\varepsilon_{2}$ in the Ref. [1], we consider a more general case,where $\varepsilon_{1}=\varepsilon_{2}$ is not strictly required in our protocol, and obtain the condition that $\varepsilon_{1}, \varepsilon_{2}$  should fulfill to generate new randomness (the choices of states and measurements are derived from $\varepsilon_{1}$-$free$ source and $\varepsilon_{2}$-$free$ source, respectively). A new 2-dimensional quantum witness is gained in the process of randomness certification. Furthermore, the analytic  relationship between the generated randomness and the 2-dimensional quantum witness violation is acquired.

This paper is structured as follows. In Sec. \uppercase\expandafter{\romannumeral 2}, we recall the definition of partially free random sources and introduce a SDI randomness expansion protocol with partially free sources. In Sec. \uppercase\expandafter{\romannumeral 3}, the condition which partially free sources should satisfy to generate new randomness, and certification parameters are obtained. In Sec. \uppercase\expandafter{\romannumeral 4}, the analytic  relationship between the generated randomness and 2-dimensional quantum witness violation is concluded. In Sec. \uppercase\expandafter{\romannumeral 5}, we summarize our results.

\section{\label{sec:level1}model description}

\theoremstyle{remark}
\newtheorem{definition}{\indent Definition}
\newtheorem{lemma}{\indent Lemma}
\newtheorem{theorem}{\indent Theorem}
\newtheorem{corollary}{\indent Corollary}

\def\QEDclosed{\mbox{\rule[0pt]{1.3ex}{1.3ex}}}
\def\QED{\QEDclosed}
\def\proof{\indent{\em Proof}.}
\def\endproof{\hspace*{\fill}~\QED\par\endtrivlist\unskip}

In order to better explain our theory, first of all, we give a detailed definition of partially free random sources in this section.

Let $X$ be a variable, considering its causal structure in relativistic space time, we call a variable $\lambda$ cannot be caused by $X$ if $\lambda$ are not in the future lightcone of X. Denote the parameter $ \Lambda$ as the set of variables which cannot be caused by $X$ and are interested in our devices. The variables in $ \Lambda$  may be provided by an eavesdropper or a higher theory [1].

\begin{definition}
  A variable bit $X$ is called $\varepsilon$-$free$ bit, $\varepsilon<\frac{1}{2}$, if it satisfies $|P(0|\Lambda=\lambda)-\frac{1}{2}|\leq\varepsilon$ for all $\lambda\in \Lambda$. Particularly, $X$ is called \emph{perfectly  free} bit as $\varepsilon=0$.
\end{definition}
 In this paper, we say that bits are picked according to $\varepsilon$-$free$ source if each bit is $\varepsilon$-$free$ and independents of other bits.

Secondly, we introduce a SDI randomness expansion protocol with partially free sources based on $n\rightarrow1$ quantum random access codes (QRACs) (see Fig 1). Based on the typical causal structure of our protocol [1], we assume that $\lambda$ may be correlated with two sources of weak randomness, the states prepared by Alice and the measurements performed by Bob.

 A detailed description of our scenario is described as follows: Alice picks $n$ bits $a=a_{0}a_{1}\cdot\cdot\cdot a_{n-1}$ according to $\varepsilon_{1}$-$free$ source $S_{1}$ and encoded to 1 qubit $\rho_{a,\lambda}$, then Alice sends it to Bob via quantum channel. Bob performs two dimensional measurement $\{M_{y,\lambda}^{b},b=0,1\}$ decided by $y=0,1,\cdot\cdot\cdot, n-1$ which is picked according to  $\varepsilon_{2}$-$free$ source $S_{2}$, and emits the measurement outcome $b$. In particular, there is not entanglement in the devices.

\begin{figure}
\includegraphics[width=0.5\textwidth]{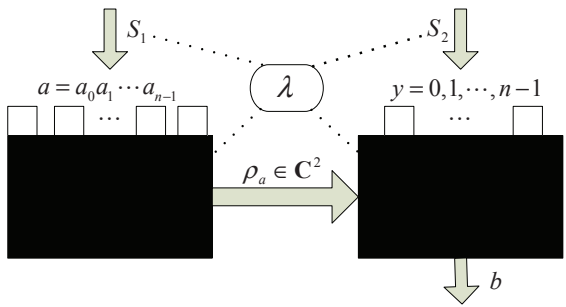}
\caption{SDI randomness expansion with partially free sources. The dashed line represent that the hidden variable $\lambda$ may be correlated with these parts. Our protocol consists of two black box, which do not contain entanglement, in safe area.}
\end{figure}

In this paper, we construct the 2-dimensional quantum witness using the expected success probability which is different from the Ref. [11] and draws better conclusions.

\emph{The expected success probability} for the scenario is
\begin{equation}
E\equiv\sum_{a,y}P(a,y)P(b=a_{y}|a,y)=\sum_{\lambda}P(\lambda)E_{\lambda},
\end{equation}
where $E_{\lambda}=\sum_{a,y}P(a,y|\lambda)P(b=a_{y}|a,y,\lambda)$ and $P(b|a,y,\lambda)=$tr$(\rho_{a,\lambda}M_{y,\lambda}^{b})$.

Probability distribution of $P(a,y,b)$ can be estimate by repeating the procedure many times, the value of $E$ can then be estimated.

Thirdly, we introduce the definition of the min-entropy function:
\begin{equation}
H_{\infty}(B|A,Y,\Lambda)\equiv -\log_{2}\max_{a,y,b,\lambda}
\sum_{\lambda\in\Lambda}P(\lambda)P(b|a,y,\lambda)
\end{equation}
to quantify the randomness of the measurement outcome for the scenario with the set $\Lambda$.

Here the SDI randomness expansion with partially free sources based on $2\rightarrow 1$ QRAC is primarily discussed. The feasible region and the randomness certification of our protocol are explored in next section.

\section{Feasible Region and randomness certification}

In the DI randomness amplification proposed by R. Colbeck \emph{et al.} [1], only one case of $\varepsilon_{1}=\varepsilon_{2}$ is discussed, and the relationship between quantum dimension witness and the min-entropy bound cannot be given as there are infinite parameters needed to be considered.
In this section, we relax the assumption of $\varepsilon_{1}=\varepsilon_{2}$. Namely, the random resources of Alice is actually not required to be same as Bob's, and obtain the feasible region. The good partially free sources are quite precious resource, our setting benefits to allocate partially free sources more reasonably and effectively.
On the other aspect, the figure of the relationship between 2-dimensional quantum witness violation and the min-entropy bound will be obtained through an optimization process.

\begin{definition}
If there exists a  protocol about SDI randomness expansion with partially free sources where Alice and Bob have the $\varepsilon_{i}$-$free$ source $S_{i},i=0,1$, respectively, and new randomness is certified, the pair $(\varepsilon_{1},\varepsilon_{2})$ is called a feasible pair. \emph{The Feasible Region} \textbf{$R$} of SDI randomness expansion with partially free sources is the set of all feasible pairs $(\varepsilon_{1},\varepsilon_{2})$.
\end{definition}

It is evident for any $\lambda\in\Lambda$ that the randomness extracted from the outcome $b$ will reduce to 0 with the increase of the distance between the probability distribution of $P(a,y|\lambda)$ and the uniform distribution on $a,y$. We assume that the eavesdropper attacks our devices in order to make our protocol get the least randomness and  attempt not to led us finding that the random sources has been changed. To achieve his targets, the eavesdropper has to let

\begin{eqnarray}
|P(a_{i}=0|\lambda)-\frac{1}{2}|&=&\varepsilon_{1},i=0,1.\nonumber\\
|P(y=0|\lambda)-\frac{1}{2}|&=&\varepsilon_{2},
\end{eqnarray}
for any $\lambda\in\Lambda$ and

\begin{equation}
P(a,y)=\sum_{\lambda}P(\lambda)P(a,y|\lambda)=\frac{1}{8}
\end{equation}
 for any $a\in\{00,01,10,11\}, y\in\{0,1\}.$

 Without loss of generality, we can assume that there are only 8 hidden variables $\lambda_{k}, k=0,1,\cdot\cdot\cdot,7$ corresponding to 8 cases in Eq. (3). For the sake of convenience, let
 \begin{eqnarray}
P(a_{i}=0|\lambda_{k})&=&\frac{1}{2}+(-1)^{k_{i}}\varepsilon_{1},i=0,1.\nonumber\\
P(y=0|\lambda)&=&\mbox{}\frac{1}{2}+(-1)^{k_{2}}\varepsilon_{2}.
\end{eqnarray}
where $k_{0}k_{1}k_{2}$ is the binary notation of $k$. It is easy to see that the eavesdropper can achieve Eq. (3) and Eq. (4) at the same time, see appendix A for the proof.

Under the above attack of the eavesdropper, if for a pair $(\varepsilon_{1},\varepsilon_{2})$, 2-dimensional quantum witness violation still exist and the min-entropy is larger than $0$ as  2-dimensional quantum witness violation reach its maximum, then we can say that the pair $(\varepsilon_{1},\varepsilon_{2})$ belongs to the feasible region  \textbf{$R$}.

Denote $E_{\lambda_{k}, c}$ as the expected success probability with parameter $\lambda_{k}$ through a classical process. For any $k$, the maximum value of  $E_{\lambda_{k}, c}$ can be obtained using the encoding map $a_{0}a_{1}\rightarrow a_{k_{2}}$, the bit $a_{k_{2}}$ decoding map $0\rightarrow0, 1\rightarrow1$ and the bit $a_{(1-k_{2})}$ decoding map $0,1\rightarrow k_{(1-k_{2})}$, and $E_{\lambda_{k}, c}$ reach the same maximum value for any $k$. Obviously, the maximum value of $E$ through a classical process is
\begin{eqnarray}
E_{c}=\frac{3}{4}+\frac{1}{2}(\varepsilon_{1}+\varepsilon_{2})-\varepsilon_{1}\varepsilon_{2}.
\end{eqnarray}

Denote $E_{\lambda_{k}, q}$ as the expected success probability with parameter $\lambda_{k}$ through a quantum process, $E_{\lambda_{k}}$ apparently meet the linear relationship. For any $k$, to reach the maximum value of $E_{\lambda_{k}, q}$, as a general rule, every quantum state $\rho_{a,\lambda_{k}}$ and positive operator valued measure (POVM) $\{M_{y,\lambda}^{0},M_{y,\lambda}^{1}\}$ performed in the 2-dimensional space should be considered.
Here, each mix state can be written as a convex combination of pure states.
On the other hand, any POVM can be described  as a convex combination of  projective measurements, which include projective measurements with rank 1, measurements $ \{I, 0\}$ and $\{0, I\}$ [12]. Different from the Ref. [8, 9], we pinpoint that measurements $ \{I, 0\}$ and $\{0, I\}$ also need to be considered in our protocol.
 Nevertheless, we can prove that once measurements $ \{I, 0\}$ or $\{0, I\}$ are chosen, the relationship  $E_{\lambda_{k}, q}\leq E_{c}$  will be satisfied and is tight. In conclusion, here only pure states and projective measurements with rank 1 need to considered.

To visualize the pure states and projective measurements with rank 1, we consider Bloch sphere representation. Without loss of generality, set the Bloch sphere representation of measurement $\{M_{y,\lambda_{k}}^{0},M_{y,\lambda_{k}}^{1}\}$ as  $\{\textbf{\emph{v}}_{y,\lambda_{k}},-\textbf{\emph{v}}_{y,\lambda_{k}}\}$ and $\textbf{\emph{v}}_{0,\lambda_{k}}=(1,0,0)$ for any $k$. Let the Bloch vector $\textbf{\emph{r}}_{a,\lambda_{k}}$ be the Bloch sphere representation of the pure state $\rho_{a,\lambda_{k}}$.

For any pair $(\varepsilon_{1},\varepsilon_{2})$, define
\begin{equation}
t \equiv\frac{8\varepsilon_{1}^{2}(1+4\varepsilon_{2}^{2})}{1+16\varepsilon_{1}^{4}-4\varepsilon_{2}^{2}-64\varepsilon_{1}^{4}\varepsilon_{2}^{2}}\geq0.
\end{equation}
\begin{equation}
\textbf{\emph{v}}_{a,\lambda_{k}}\equiv\sum_{i=0,1}(-1)^{a_{i}}(\frac{1}{2}+(-1)^{k_{2}}\varepsilon_{2})\textbf{\emph{v}}_{i,\lambda_{k}}.
\end{equation}

If $t>1$, the optimal encoding-decoding strategy for any $k$: $\textbf{\emph{v}}_{1,\lambda_{k}}=(1, 0, 0)$ and
$\textbf{\emph{r}}_{a,\lambda_{k}}=\textbf{\emph{v}}_{a,\lambda_{k}}/\|\textbf{\emph{v}}_{a,\lambda_{k}}\|$. Hence the maximum value of $E$ is
 \begin{equation}
E=\frac{3}{4}+\frac{1}{2}\varepsilon_{2}+\varepsilon_{1}^{2}(1-2\varepsilon_{2})\leq E_{c}.
 \end{equation}

If $t\leq1$, the optimal encoding-decoding strategy for any $k$: $\textbf{\emph{v}}_{1,\lambda_{k}}=((-1)^{k_{0}+k_{1}}t,\sqrt{1-t^{2}}, 0)$ and
$\textbf{\emph{r}}_{a,\lambda_{k}}=\textbf{\emph{v}}_{a,\lambda_{k}}/\|\textbf{\emph{v}}_{a,\lambda_{k}}\|$.
After the simple analysis and calculation, $E_{\lambda_{k},q}$ will reach the same maximum value and the maximum value of $E$ through a quantum process is
\begin{eqnarray}
E_{q}=\frac{1}{2}+\frac{1}{2}\sqrt{\frac{1}{2}+8\varepsilon_{1}^{4}+2\varepsilon_{2}^{2}+32\varepsilon_{1}^{4}\varepsilon_{2}^{2}}.
\end{eqnarray}

For $E=E_{q}$, the min-entropy  is
\begin{eqnarray}
H_{\infty}(B|A,Y,\Lambda)=1-\log_{2}(1+\frac{t+\delta}{\sqrt{\delta^{2}+2t\delta+1}}),
\end{eqnarray}
where $\delta=(1+2\varepsilon_{2})/(1-2\varepsilon_{2}).$

This implies that a pair $(\varepsilon_{1},\varepsilon_{2})$ belongs to the feasible region \textbf{$R$} once it satisfies $t\leq1$ , $ E_{ c}<E_{q}$ and $H_{\infty}(B|A,Y,\Lambda)>0$. Then the feasible region is obtained and demonstrated in Fig 2. Moreover, a new tight bound for 2-dimensional classical and quantum systems are given as $E_{c}$  and $E_{q}$, respectively. Namely, a new 2-dimensional quantum witness is presented.
See appendix B for a detailed calculation of Eq. (9)-(11).

\begin{figure}
\includegraphics[width=0.5\textwidth]{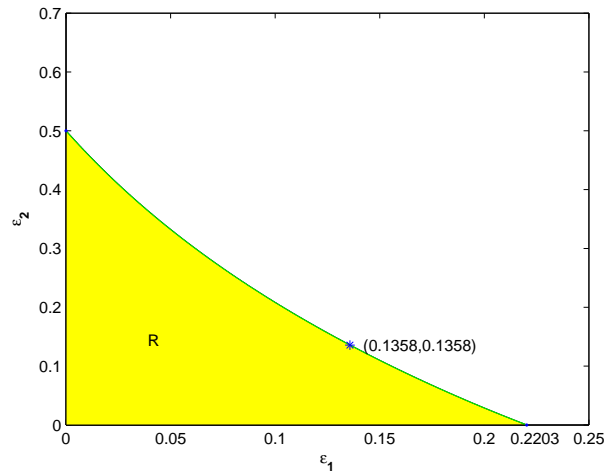}
\caption{The feasible region \textbf{$R$} of SDI randomness expansion with partially free random sources is the yellow region under the green line but does not include the green line.  Alice pick $a$ according to  $\varepsilon_{1}$-$free$ source $S_{1}$ and  Bob pick $y$ according to  $\varepsilon_{2}$-$free$ source $S_{2}$, respectively.}
\end{figure}

Next,  for arbitrary pair  $(\varepsilon_{1},\varepsilon_{2})\in$ \textbf{$R$}, we begin to discuss  the min-entropy bound for a given  expected success probability $E$, which can be resolved by the following optimization problem:
\begin{eqnarray}
& &\min_{a,y,b,\lambda}H(B|A,Y,\Lambda)\nonumber\\
& &\mbox{}\text{subject to}: E=\sum_{k=0}^{7}P(\lambda_{k})E_{\lambda_{k}},\nonumber\\
& &\mbox{}E_{\lambda_{k}}=\sum_{a,y}P(a,y|\lambda_{k})P(b=a_{y}|a,y,\lambda_{k}),
\end{eqnarray}
the optimization is carried out by quantum states $\rho_{a,\lambda}$ and POVMs $\{M_{y,\lambda}^{0},M_{y,\lambda}^{1}\}$ chosen in 2-dimensional Hilbert space for $a\in\{00,01,10,11\}$ and $y\in\{0,1\}$.

After that we can estimate the min-entropy bound. Then true random numbers can be produced by a randomness extractor [13]. In fact, it plays an important role on many aspects that the min-entropy bound can be estimated as the analytic function of 2-dimensional quantum witness violation, such as security analysis of SDI randomness expansion [14].

\section{analytic function}

In the SDI randomness expansion proposed by Li \emph{et al.}, the figure of the relationship between 2-dimensional quantum witness and the min-entropy is given, but the analytic relationship is not discussed. In this section, for arbitrary pair  $(\varepsilon_{1},\varepsilon_{2})\in$ \textbf{$R$}, we explore the analytic relationship between 2-dimensional quantum witness
\begin{equation}
E=\sum_{k=0}^{7}P(\lambda_{k})E_{\lambda_{k}}
\end{equation}
and the min-entropy bound $H(B|A,Y,\Lambda)=-\log_{2}p$, where
\begin{equation}
1/2+(t+\delta)/(2\sqrt{\delta^{2}+2t\delta+1})\leq p\leq 1\nonumber
\end{equation}
deduced from Eq. (11).

We might take $\lambda_{0}$ as example. To depict a  encoding-decoding strategy influenced by the hidden variable $\lambda_{0}$, we extract two parameters $(E_{\lambda_{0}},\max_{a,y,b}P(b|a,y,\lambda_{0}))$, which can be regarded as points in the 2-dimensional coordinate system.

For a given encoding-decoding strategy, $E_{\lambda_{0}}$ can be said as the convex combination of success conditional expected success probabilities obtained by pure states and projective measurements,
and $\max_{a,y,b}P(b|a,y,\lambda_{0})$ is not more than the convex combination of the maximal guess probability obtained by the same pure states and projective measurements. That is, for a given value of $E_{\lambda_{0}}$, the convex set composed of the realizable points achieved by pure states and projective measurements will provide a upper concave bound for $\max_{a,y,b}P(b|a,y,\lambda_{0})$, denote the upper bound as $p_{\lambda_{0}}$. We only discuss this upper concave bound in the following.

Apparently, $p_{\lambda_{0}}$ can be viewed as a concave function of $E_{\lambda_{0}}$. Denote $p_{\lambda_{0}}=C(E_{\lambda_{0}})$, $C$ is a concave function. On the other hands, with the increase of $E_{\lambda_{0}}$, the randomness generated by the quantum process will also be monotone increasing, as consequences $C$ is a continuous and decreasing function.

 Fortunately, this discussion  also applies to other hidden variables $\lambda_{k}, k\neq0.$ For any realizable point $(E_{\lambda_{0}},\max_{a,y,b}P(b|a,y,\lambda_{0}))$, other hidden variables can realize through a single code. It is to say that other hidden variables will reach the same bound  $p_{\lambda_{k}}$ as $p_{\lambda_{0}}$ for  $E_{\lambda_{k}}=E_{\lambda_{0}},$  i.e., $p_{\lambda_{k}}=C(E_{\lambda_{k}})$.

For the given $E$ as indicated in Eq. (13), the lower bound of the min-entropy is
\begin{eqnarray}
H(B|A,Y,\Lambda)&=&-\log_{2}\sum_{k=0}^{7}P(\lambda_{k})\max_{a,y,b,\lambda_{k} }P(b|a,y,\lambda_{k})\nonumber\\
&=&\mbox{}-\log_{2}\sum_{k=0}^{7}P(\lambda_{k})P_{\lambda_{k}}.
\end{eqnarray}
Using the Jensen's inequality, it is natural that if and only if $E_{\lambda_{k}}=E_{\lambda_{k'}}, k\neq k'$ , the lower bound of min-entropy will be reached.
Without loss of generality, we might take $E=E_{\lambda_{0}}$,  then $p=p_{\lambda_{0}}$.  The lower bound of min-entropy can be described as
\begin{equation}
H(B|A,Y,\Lambda)=-\log_{2}C(E).
\end{equation}

The next work mainly describe the function $C$. Denote $E_{l}$  as
 \begin{equation}
 E_{l}=\max\{E_{\lambda_{0}}:C(E_{\lambda_{0}})=1\}.
 \end{equation}

 Obviously, we can deduce that $C(E)=1$ as $E\leq E_{l}$ according to the  monotonicity of the function $C$. Therefore, the function $C$ can be completely depicted once the one in the closed interval $[ E_{l}, E_{q}]$ is obtained and the value of $E_{l}$ is determined, which only are discussed in the following.

 Based on the Refs. [8-10], the function $C$ is monotropic as $E\in [ E_{l}, E_{q}]$. Then the function  $C$ has a inverse function denoted as $C^{-1}$, i.e.,
$E=C^{-1}(p)$. The function $C^{-1}$ can be obtained by the following  optimization:
\begin{eqnarray}
& &\max_{a,y, b, \lambda_{0}}E=\sum_{a,y}P(a,y|\lambda_{0})P(b=a_{y}|a,y,\lambda_{0})\nonumber\\
& &\mbox{}\text{subject to}: \max_{a,y,b,\lambda_{0}}P(b|a,y,\lambda_{0})=p,
\end{eqnarray}
where $1/2+(t+\delta)/(2\sqrt{\delta^{2}+2t\delta+1})\leq p\leq 1$, the optimization is carried out pure quantum states $\rho_{a,\lambda}$ and projective measurements chosen in 2-dimensional Hilbert space for $a\in\{00,01,10,11\}, y\in\{0,1\}$ and $\lambda\in\Lambda$.

Firstly, we focus on the value of $E$ achieved by arbitrary  pure states and projective measurements with rank 1 in the optimization (17). Fortunately, $E$ can be viewed as a function $G(\varepsilon_{1},\varepsilon_{2},p)$ determined by $\varepsilon_{1},\varepsilon_{2}$ and $p$ (see the appendix C).

Secondly, consider the optimization (17) achieved by arbitrary pure states and measurements $\{I, 0\}$ , $\{0, I\}$ in optimization (17), only $E=E_{c}$ as $p=1$ is obtained.

 If $E_{\varepsilon_{1},\varepsilon_{2}}\geq  E_{ c}$ is established. The strategy with measurements $\{I, 0\}$ or $\{0, I\}$  can be simulated by one with pure states and projective measurements with rank 1, where $E_{\varepsilon_{1},\varepsilon_{2}}=G(\varepsilon_{1},\varepsilon_{2},1)$. Here $E=G(\varepsilon_{1},\varepsilon_{2},p)$.

On the contrary, if $E_{\varepsilon_{1},\varepsilon_{2}}< E_{c}$, then we have to discuss the convex set which is composed of points $(G(\varepsilon_{1},\varepsilon_{2},p),p)$ and $(E_{c},1)$  to obtain the upper bound of $E$ denoted as $F(\varepsilon_{1},\varepsilon_{2},p)$, which is a function of $\varepsilon_{1},\varepsilon_{2}$ and $p$.  In fact, the points $(F(\varepsilon_{1},\varepsilon_{2},p), p)$ also provide a lower bound of min-entropy. However, whether the bound is tight or not cannot be determined.

Based on the above analysis,
 we have $E_{l}=\max\{E_{c}, E_{\varepsilon_{1},\varepsilon_{2}}\}$ and for $E_{l}\leq E\leq E_{q}$,
  $$ E=C^{-1}(p)=\left\{
 \begin{array}{lcl}
 G(\varepsilon_{1},\varepsilon_{2},p), &    &if \ E_{\varepsilon_{1},\varepsilon_{2}}\geq E_{c},\\
 F(\varepsilon_{1},\varepsilon_{2},p)&    &if \ E_{\varepsilon_{1},\varepsilon_{2}}< E_{c},
 \end{array}
 \right.$$
  where $1/2+(t+\delta)/(2\sqrt{\delta^{2}+2t\delta+1})\leq p\leq 1$.

In fact, the function $G$ and $F$ can describe the relationship between the min-entropy bound and the 2-dimensional quantum witness violation in detail.

Denote $\beta=\arccos(2p-1)$, we have
\begin{eqnarray}
G(\varepsilon_{1},\varepsilon_{2},p)=\max_{\alpha\in[0,\pi-4\beta], i=1,2}\{G_{i}(\varepsilon_{1},\varepsilon_{2},p,\alpha)\}.\nonumber
\end{eqnarray}
The analytic functions $G_{1},G_{2}$ are
\begin{eqnarray}
G_{1}(\varepsilon_{1},\varepsilon_{2},p,\alpha)&=&\frac{1}{2}+\frac{1}{2}(\frac{1}{2}-\varepsilon_{1})^{2}(\frac{1}{2}-\varepsilon_{2})[\delta \cos\beta\nonumber\\
& &\mbox{}+\cos(\beta+\alpha)+f(\varepsilon_{1},\varepsilon_{2},p,\alpha)].
\end{eqnarray}
\begin{eqnarray}
G_{2}(\varepsilon_{1},\varepsilon_{2},p,\alpha)&=&\frac{1}{2}+\frac{1}{2}(\frac{1}{2}-\varepsilon_{1})^{2}(\frac{1}{2}-\varepsilon_{2})[\delta \sigma \cos\beta\nonumber\\
& &\mbox{}+\sigma \cos(\beta+\alpha)+g(\varepsilon_{1},\varepsilon_{2},p,\alpha)],\nonumber
\end{eqnarray}
where $\sigma=(1+2\varepsilon_{1})/(1-2\varepsilon_{1})$ ($f,g$ see Eq. (C12), Eq. (C13), respectively).

The function $F$ can be depicted by the function $G$ and
\begin{eqnarray}
F(\varepsilon_{1},\varepsilon_{2},p)&=&(G(\varepsilon_{1},\varepsilon_{2},p_{0})-E_{c})(1-p)/(1-p_{0})\nonumber\\
& &\mbox{}+E_{c}
\end{eqnarray}
as $p\geq p_{0}$,
$F(\varepsilon_{1},\varepsilon_{2},p)=G(\varepsilon_{1},\varepsilon_{2},p)$
as $p<p_{0}$,
 where $p_{0}$ satisfies
\begin{equation}
\frac{G(\varepsilon_{1},\varepsilon_{2},p_{0})-E_{c}}{p_{0}-1}=\min_{p}\{\frac{G(\varepsilon_{1},\varepsilon_{2},p)-E_{c}}{p-1}\}.\nonumber
\end{equation}

In particular, for the cases of $\varepsilon_{1}=\varepsilon_{2}<0.1358$; $\varepsilon_{1}<0.2203, \varepsilon_{2}=0$ and $\varepsilon_{1}=0, \varepsilon_{2}<0.5$,  we have
\begin{center}
$E_{\varepsilon_{1},\varepsilon_{2}}=\max_{\alpha\in[0,\pi-4\beta]}\{G_{1}(\varepsilon_{1},\varepsilon_{2},1,\alpha)\}.$
\end{center}
In the Ref. [8], the analytic relationship between the min-entropy bound and the 2-dimensional quantum witness violation is $E=G(0,0,2^{-H_{\infty}(B|A,Y,\Lambda)})$.

\begin{figure}
\includegraphics[width=0.5\textwidth]{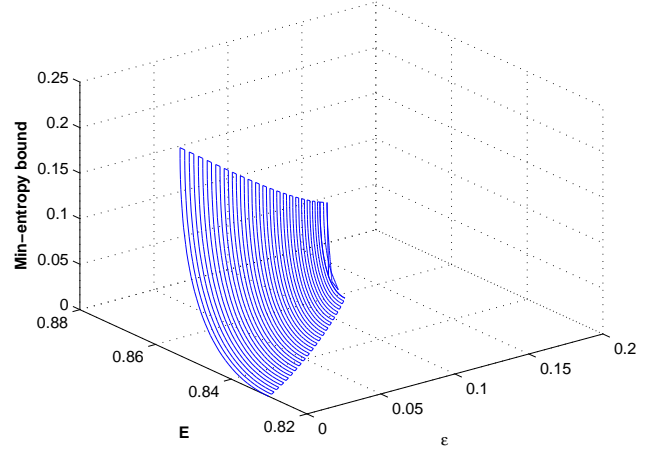}
\caption{The relationship between the min-entropy bound and the 2-dimensional quantum witness for $0\leq\varepsilon<0.1358$, where the choices of states and measurements are derived from $\varepsilon$-$free$ sources $S_{1}$ and $S_{2}$, respectively.}
\end{figure}

Many works only consider projective measurements with rank 1 just because they happen to satisfy $E_{\varepsilon_{1},\varepsilon_{2}}\geq  E_{c}$, such as the Refs. [8, 9]. We might take $\varepsilon_{1}=\varepsilon_{2}=\varepsilon$ as example. In fact, $E_{\varepsilon_{1},\varepsilon_{2}}\geq  E_{c}$ as $\varepsilon\leq 0.12348$. But $E_{\varepsilon_{1},\varepsilon_{2}}< E_{c}$ as $0.12348<\varepsilon <0.1358$. This shows that the situation of $E_{\varepsilon_{1},\varepsilon_{2}}<E_{c}$ will occur with the increase of $\varepsilon_{1},\varepsilon_{2}$ and the measurements $\{I,0\}$  and $\{0.I\}$  must be taken into consideration.
Furthermore, the relationship between the min-entropy bound and the 2-dimensional quantum witness for  $\varepsilon_{1}=\varepsilon_{2}$ is demonstrated as the Fig.3.
\section{Conclusion}

We proved that partially free sources can bring us new randomness and proposed a SDI randomness expansion protocol with partially free sources based on $2\rightarrow1$ QRAC. In our protocol, the condition that the partially free sources should satisfy to generate new randomness was gained without strictly requiring $\varepsilon_{1}=\varepsilon_{2}$ (the choices of states and measurements are derived from $\varepsilon_{1}$-$free$ source and $\varepsilon_{2}$-$free$ source, respectively). Furthermore, a new 2-dimensional quantum witness and the analytic relationship between the generated randomness and the 2-dimensional quantum witness violation were obtained. In addition, the advantage of no containing entanglement which is introduced in the Ref. [8, 9] also apply to our protocol. We conjecture that it can get better results in the SDI randomness expansion with partially free sources based on $n\rightarrow1$ QRACs for $n\geq3$.

\begin{acknowledgments}
The authors would like to thank Z. Q. Yin for many valuable suggestions and Q. N. Zhou for Fig.3. This work is supported by NSFC (Grant Nos. 61272057, 61170270,U1304604), Beijing Higher Education Young Elite Teacher Project (Grant Nos. YETP0475, YETP0477).
\end{acknowledgments}

\appendix
\section{}

The assumption in Eq. (5) apparently satisfy Eq. (3). In addition, Eq. (4) is equivalent to
\begin{eqnarray}
P(a_{i}=0)&=&\sum_{k=0}^{7}P(\lambda_{k})P(a_{0}=0|\lambda_{k})=\frac{1}{2},i=0,1.\nonumber\\
P(y=0)&=&\sum_{k=0}^{7}P(\lambda_{k})P(y=0|\lambda_{k})=\frac{1}{2}.
\end{eqnarray}

The Eq. (A1) can be written as
\begin{eqnarray}
\sum_{k=0,1,2,3}P(\lambda_{k})&-&\sum_{k=4,5,6,7}P(\lambda_{k})=0.\nonumber\\
\sum_{k=0,1,4,5}P(\lambda_{k})&-&\sum_{k=2,3,6,7}P(\lambda_{k})=0.\nonumber\\
\sum_{k=0,2,4,6}P(\lambda_{k})&-&\sum_{k=1,3,5,7}P(\lambda_{k})=0.
\end{eqnarray}
They are the linear equations of $\lambda_{k}, k=0,1,\cdot\cdot\cdot,7$. It is easy to see there  must be many solutions to Eq. (A2) as there are three equations but eight variables.
\section{}
The expected success probability with variable $\lambda_{k}$ is
\begin{equation}
 E_{\lambda_{k}}=\sum_{a,y}P(a,y|\lambda_{k})P(b=a_{y}|a,y,\lambda_{k})
\end{equation}
for $k=0,1,\cdot\cdot\cdot,7.$

We might take the $\lambda_{0}$ as example. Set the Bloch sphere representation of the pure state $\rho_{a,\lambda_{0}}$ , projective measure $\{M_{y,\lambda_{0}}^{0},M_{y,\lambda_{0}}^{1}\}$ as the Bloch vector $\textbf{\emph{r}}_{a,\lambda_{0}}$, $\{\textbf{\emph{v}}_{y,\lambda_{0}},-\textbf{\emph{v}}_{y,\lambda_{0}}\}$ for $a\in\{00,01,10,11\}$ and $y\in\{0,1\}$, respectively. Without loss of generality, let $\textbf{\emph{v}}_{0,\lambda_{0}}=(1,0,0).$ By the Ref. [10], we can know
\begin{eqnarray}
P(b|a,y,\lambda_{0})&=&tr(\rho_{a,\lambda_{0}}M_{y,\lambda_{0}}^{b})\nonumber\\
&=&\mbox{}\frac{1}{2}(1+\textbf{\emph{r}}_{a,\lambda_{0}}\cdot(-1)^{b}\textbf{\emph{v}}_{y,\lambda_{0}}),
\end{eqnarray}
where $``\cdot"$ denotes the \emph{inner product}.

With a small amount of calculation, we get
\begin{eqnarray}
E_{\lambda_{0}}&=&\frac{1}{2}+\frac{1}{2}\sum_{a,y}P(a,y|\lambda_{0})\textbf{\emph{r}}_{a,\lambda_{0}}\cdot(-1)^{b}\textbf{\emph{v}}_{y,\lambda_{0}}\nonumber\\
&=&\mbox{}\frac{1}{2}+\frac{1}{2}\sum_{a}P(a|\lambda_{0})\textbf{\emph{r}}_{a,\lambda_{0}}\cdot\textbf{\emph{v}}_{a,\lambda_{0}}\nonumber\\
&\leq&\mbox{}\frac{1}{2}+\frac{1}{2}\sum_{a}P(a|\lambda_{0})\|\textbf{\emph{v}}_{a,\lambda_{0}}\|\nonumber\\
\end{eqnarray}
where $\textbf{\emph{v}}_{a,\lambda_{0}}$ defined as Eq. (8). Only the case of $\parallel\textbf{\emph{v}}_{a,\lambda_{0}}\parallel\neq 0$ is discussed in the following. If and only if $\textbf{\emph{r}}_{a,\lambda_{0}}=\textbf{\emph{v}}_{a,\lambda_{0}}/\|\textbf{\emph{v}}_{a,\lambda_{0}}\|$, the Eq. (B3) can achieve the maximum value.

Furthermore, set $\theta$ is the angle between $\textbf{\emph{v}}_{0,\lambda_{0}}$ and $\textbf{\emph{v}}_{1,\lambda_{0}}$, then
\begin{eqnarray}
& &\|\textbf{\emph{v}}_{00,\lambda_{0}}\|^{2}+\|\textbf{\emph{v}}_{01,\lambda_{0}}\|^{2}=1+4\varepsilon_{2}^{2}.\nonumber\\
& &\mbox{}P(00|\lambda_{0})+P(11|\lambda_{0})=\frac{1}{2}+2\varepsilon_{1}^{2}.\nonumber\\
& &\mbox{}P(01|\lambda_{0})+P(10|\lambda_{0})=\frac{1}{2}-2\varepsilon_{1}^{2},
\end{eqnarray}
where  Alice and Bob have the $\varepsilon_{i}$-$free$ source $S_{i},i=0,1$ to pick $a,y$, respectively.

With the knowledge of Eq. (B4), we have
\begin{eqnarray}
 \sum_{a}P(a|\lambda_{0})\|\textbf{\emph{v}}_{a,\lambda_{0}}\|&=&(\frac{1}{2}+2\varepsilon_{1}^{2})\|\textbf{\emph{v}}_{00,\lambda_{0}}\|\nonumber\\
& &\mbox+(\frac{1}{2}-2\varepsilon_{1}^{2})\|\textbf{\emph{v}}_{01,\lambda_{0}}\|\nonumber\\
&\leq&\mbox{}\sqrt{\frac{1}{2}+8\varepsilon_{1}^{4}}\sqrt{1+4\varepsilon_{2}^{2}}.
\end{eqnarray}
 If and only if $\|\textbf{\emph{v}}_{00,\lambda_{0}}\|/\|\textbf{\emph{v}}_{01,\lambda_{0}}\|=(1+4\varepsilon_{1}^{2})/(1-4\varepsilon_{1}^{2})$, that is, the angle $\theta$ must satisfies $\cos\theta=t$, $t$ is defined as Eq. (7), the Eq. (B5) can achieve the maximum value.

 If $t\leq1$ for a pair $(\varepsilon_{1},\varepsilon_{2})$. Fortunately, let $\theta=\arccos t$, we will reach the maximum value of $E_{\lambda_{0}}$ denoted as
\begin{eqnarray}
E_{\lambda_{0}, q}^{ max}=\frac{1}{2}+\frac{1}{2}\sqrt{\frac{1}{2}+8\varepsilon_{1}^{4}+2\varepsilon_{2}^{2}+32\varepsilon_{1}^{4}\varepsilon_{2}^{2}}.
\end{eqnarray}
At the same time, the maximum success probability with variable $\lambda_{0}$ can be obtained:
\begin{eqnarray}
P(b=0|00,0,\lambda_{0})=\frac{1}{2}(1+\frac{t+\delta}{\sqrt{\delta^{2}+2t\delta +1}}).
\end{eqnarray}

If $t>1$ for a pair $(\varepsilon_{1},\varepsilon_{2})$, we have to set $\cos\theta=1$, i.e., $\theta=0$  to obtain the maximum value
 \begin{eqnarray}
E_{\lambda_{0}, q}^{ max}=\frac{3}{4}+\frac{1}{2}\varepsilon_{2}+\varepsilon_{1}^{2}(1-2\varepsilon_{2}).
\end{eqnarray}

\section{}

 For the sake of simplicity, suppose $E=E_{\lambda_{0}}$ and $p=p_{\lambda_{0}}$.

 To reach the maximum value of $E$ satisfying a condition that the probability distribution satisfies $\max_{a,y,b}P(b|a,y,\lambda_{0})=p=\frac{1}{2}(1+\cos\beta)$, we consider the question that which one of 16 probability  $P(b|a,y,\lambda_{0})$ should reach $p$, where

\begin{equation}
 (t+\delta)/\sqrt{\delta^{2}+2t\delta  +1}\leq\cos\beta\leq 1.
\end{equation}

 Apparently, it must be a guessing success probability and need to satisfy the following conditions:

  ($\romannumeral1$) It must be the probability $P(b=a_{0}|a,y=0,\lambda_{0})$ for $a\in\{00,01,10,11\}$.  Since $P(b=a_{0}|a,y=0,\lambda_{0})\geq P(b=a_{1}|a,y=1,\lambda_{0})$ can achieve a larger value of $E$ concluded from  $P(y=0|\lambda_{0})\geq P(y=1|\lambda_{0})$.

  ($\romannumeral2$) The bolch  vectors $\textbf{\emph{r}}_{a,\lambda_{0}}$, $\textbf{\emph{v}}_{y,\lambda_{0}}$ for all $a,y$ are in a plane,  $\textbf{\emph{r}}_{a,\lambda_{0}}$ fall on the area between  $(-1)^{a_{0}}\textbf{\emph{v}}_{0,\lambda_{0}}$ and $(-1)^{a_{1}}\textbf{\emph{v}}_{1,\lambda_{0}}$.

  Then only four cases of $P(b=a_{0}|a,y=0,\lambda_{0})=p$  for $a\in\{00,01,10,11\}$ need to be discussed on the strict precondition ($\romannumeral2$). It is noteworthy that
  \begin{equation}
  P(b=a_{1}|a,y=1,\lambda_{0})\leq P(b=a_{0}|a,y=0,\lambda_{0})
  \end{equation}
  always be established no matter which one of $P(b=a_{0}|a,y=0,\lambda_{0})$ reach $p$.

 Case 1: Let
 \begin{eqnarray}
 P(b=1|11,y=0,\lambda_{0})=p,
 \end{eqnarray}
 that is, the angle between $-\textbf{\emph{v}}_{0,\lambda_{0}}$ and $\textbf{\emph{r}}_{11,\lambda_{0}}$ is $\beta$, suppose the angle between $\textbf{\emph{r}}_{11,\lambda_{0}}$ and $-\textbf{\emph{v}}_{1,\lambda_{0}}$ is $\beta+\alpha.$ Since $ P(b=1|11,y=1,\lambda_{0})\leq p$, let $ \alpha\geq 0$ . Fortunately, for  any $ \alpha\geq 0$, there is at least one choice can satisfy $P(b=0|00,y,\lambda_{0})\leq p$ which let $\textbf{\emph{r}}_{00,\lambda_{0}}$  along  the same direction as $\textbf{\emph{v}}_{0,\lambda_{0}}+\textbf{\emph{v}}_{1,\lambda_{0}}$.

 In order to ensure
 \begin{eqnarray}
 P(b=0|01,y=0,\lambda_{0})&\leq&p,\nonumber\\
 P(b=1|10,y=0,\lambda_{0})&\leq&p,
\end{eqnarray}
 then $\alpha\leq \pi -4\beta$ concluded from $[\pi-(2\beta+\alpha)]/2\geq\beta.$ We confirm that the range of $\alpha$ is $[0, \pi -4\beta]$.

   If the value of $\alpha$ is determinated, the angle between $\textbf{\emph{v}}_{0,\lambda_{0}}$ and $\textbf{\emph{v}}_{1,\lambda_{0}}$ is determined, then the vector $\textbf{\emph{v}}_{1,\lambda_{0}}$ is determined since we have set $\textbf{\emph{v}}_{0,\lambda_{0}}=(1,0,0).$

 Next only need to consider how to place $\textbf{\emph{r}}_{a,\lambda_{0}}$  for the purposes of reaching the largest $E$.

 Firstly, $\textbf{\emph{r}}_{00,\lambda_{0}}$ has been determined.

Secondly, denote $\varphi$ as the angle between  $\textbf{\emph{r}}_{01,\lambda_{0}}$ and $\textbf{\emph{v}}_{0,\lambda_{0}}$ . $\varphi\leq\pi/2$ is obtained derived from the Eq. (C2).
To get a lager value of $E$,
 we want to set
 \begin{eqnarray}
  \textbf{\emph{r}}_{01,\lambda_{0}}=\textbf{\emph{v}}_{01,\lambda_{0}}/\|\textbf{\emph{v}}_{01\lambda_{0}}\|,
   \end{eqnarray}
 but must guarantee  $\varphi \geq \beta$ for the sake of that Eq. (C4) will not be established, i.e., if Eq. (C5) want to be established, we must guarantee that
 \begin{eqnarray}
  \tan\varphi =\frac{\sin(2\beta+\alpha)}{\delta-\cos(2\beta+\alpha)}\geq  \tan\beta.
 \end{eqnarray}

Using the knowledge of trigonometric functions, the Eq. (C5) is equivalent to $\sin(3\beta+\alpha)\geq \delta\sin\beta.$ Combining the condition $\alpha\in[0, \pi -4\beta]$ and $\sin(3\beta+\alpha)\geq \delta\sin\beta$ yields  that the Eq. (C6) can be established only for $\alpha\in[a_{1}, a_{2}]$, where $a_{1}=\max \{0, \arcsin(\delta\sin\beta)-3\beta\}, a_{2}=\pi-3\beta-\arcsin(\delta\sin\beta)$.

If $\alpha\in[0, a_{1})\cup(a_{2}, \pi-4\beta],$ we have to set $\varphi=\beta$ to get a lager value of $E$.

At last, by the similar way, suppose $\textbf{\emph{r}}_{10,\lambda_{0}}=\textbf{\emph{v}}_{10,\lambda_{0}}/\|\textbf{\emph{v}}_{10\lambda_{0}}\|$ for  $\alpha\in[a_{1}, a_{2}]$. If $\alpha\in[0, a_{1})\cup(a_{2}, \pi-4\beta],$ suppose the angle between $\textbf{\emph{r}}_{10,\lambda_{0}}$ and $-\textbf{\emph{v}}_{0,\lambda_{0}}$ is $\beta$,

Assume $\textbf{\emph{a}}_{00,\lambda_{0}}=\textbf{\emph{v}}_{00,\lambda_{0}}/\|\textbf{\emph{v}}_{00,\lambda_{0}}\|$ for $\alpha\in[b_{1}, b_{2}]$. In the case of $\alpha\in[0, b_{1})\cup(b_{2}, \pi-4\beta]$, let $\beta$ the angle between  $\textbf{\emph{a}}_{00,\lambda_{0}}$ and $\textbf{\emph{v}}_{0,\lambda_{0}}$ as $\beta$, where $b_{1}=\arcsin(\delta\sin\beta)-\beta, b_{2}=\min\{\pi-4\beta, \pi-\arcsin(\delta\sin\beta)-\beta\}$.

It's worth noting that $b_{1}\leq a_{2}$ for any pair of the feasible region.

Based on the analysis of the above, we can get the analytic function of $E$ in the case 1 and denote it as $G_{1}(\varepsilon_{1},\varepsilon_{2},p,\alpha)$ shown as Eq. (20). With some calculation, we can obtain
\begin{eqnarray}
G_{1}(\varepsilon_{1},\varepsilon_{2},p,\alpha)&=&\sum_{a,y}P(a,y|\lambda_{0})P(b=a_{y}|a,y,\lambda_{0})\nonumber\\
&=&\mbox{}\frac{1}{2}+\frac{1}{2}(\frac{1}{2}-\varepsilon_{1})^{2}(\frac{1}{2}-\varepsilon_{2})[\delta \cos\beta\nonumber\\
& &\mbox{}+\cos(\beta+\alpha)+f(\varepsilon_{1},\varepsilon_{2},p,\alpha)]
\end{eqnarray}
where $f(\varepsilon_{1},\varepsilon_{2},p,\alpha)$  is displayed as Eq. (C12).

\newcounter{mytempeqncnt}
\begin{figure*}[!t]
\normalsize
\setcounter{mytempeqncnt}{\value{equation}}
\setcounter{equation}{11}

\begin{equation}
  f(\varepsilon_{1},\varepsilon_{2},p,\alpha)=\left\{
   \begin{aligned}
   & (2\delta\sigma+\sigma^{2}\delta)\cos\beta +\sigma^{2}\cos(\beta+\alpha)-2\sigma\cos(3\beta+\alpha)& & if\ \alpha\in[0,a_{1})\cup(b_{2},\pi-4\beta],\\
   & \sigma^{2}\delta\cos\beta+\sigma^{2}\cos(\beta+\alpha)+2\sigma\sqrt{\delta^{2}+1-2\delta\cos(2\beta+\alpha)}& & if\ \alpha\in[a_{1}, b_{1}),\\
   & \sigma^{2}\sqrt{\delta^{2}+1+2\delta\cos(2\beta+\alpha)}+2\sigma\sqrt{\delta^{2}+1-2\delta\cos(2\beta+\alpha)}& & if\ \alpha\in [b_{1}, a_{2}),\\
   & \sigma^{2}\sqrt{\delta^{2}+1+2\delta\cos(2\beta+\alpha)}+2\delta\sigma\cos\beta-2\sigma\cos(3\beta+\alpha)& & if\ \alpha\in [a_{2}, b_{2}].\\
   \end{aligned}
   \right.
  \end{equation}
\begin{equation}
g(\varepsilon_{1},\varepsilon_{2},p,\alpha)=\left\{
\begin{aligned}
& \delta\sigma\cos\beta+\sigma\cos(\beta+\alpha)+(\sigma^{2}\delta+\delta)\cos\beta-(\sigma^{2}+1)\cos(3\beta+\alpha)& &if \ \alpha\in[0,a_{1})\cup(b_{2},\pi-4\beta],\\
& \delta\sigma\cos\beta+\sigma\cos(\beta+\alpha)+(\sigma^{2}+1)\sqrt{\delta^{2}+1-2\delta\cos(2\beta+\alpha)}& &if \ \alpha\in[a_{1}, b_{1}),\\
& \sigma\sqrt{\delta^{2}+1+2\delta\cos(2\beta+\alpha)}+(\sigma^{2}+1)\sqrt{\delta^{2}+1-2\delta\cos(2\beta+\alpha)}& &if \ \alpha\in [b_{1}, a_{2}),\\
& \sigma\sqrt{\delta^{2}+1+2\delta\cos(2\beta+\alpha)}+(\sigma^{2}\delta+\delta)\cos\beta-(\sigma^{2}+1)\cos(3\beta+\alpha)& &if \ \alpha\in [a_{2}, b_{2}].
\end{aligned}
\right.
\end{equation}
\setcounter{equation}{\value{mytempeqncnt}}
\hrulefill
\vspace*{4pt}
\end{figure*}
Case 2: Let
 \begin{eqnarray}
 P(b=0|01,y=0,\lambda_{0})=p,\nonumber
 \end{eqnarray}
 that is, the angle between $-\textbf{\emph{v}}_{0,\lambda_{0}}$ and $\textbf{\emph{r}}_{01,\lambda_{0}}$ is $\beta$, set the angle between $\textbf{\emph{r}}_{01,\lambda_{0}}$ and $-\textbf{\emph{v}}_{1,\lambda_{0}}$ is $\beta+\alpha.$ Obviously, the range of $\alpha$ remains $[0,\pi-4\beta].$
By a similar way, denote $E$ as $G_{2}(\varepsilon_{1},\varepsilon_{2},p,\alpha)$ in this case. we have

\begin{eqnarray}
G_{2}(\varepsilon_{1},\varepsilon_{2},p,\alpha)&=&\frac{1}{2}+\frac{1}{2}(\frac{1}{2}-\varepsilon_{1})^{2}(\frac{1}{2}-\varepsilon_{2})[\delta \sigma \cos\beta\nonumber\\
& &\mbox{}+\sigma \cos(\beta+\alpha)+g(\varepsilon_{1},\varepsilon_{2},p,\alpha)]
\end{eqnarray}
The detailed description of $g$ is displayed in Eq. (C13).

Case 3: Let
 \begin{eqnarray}
 P(b=1|10,y=0,\lambda_{0})=p.\nonumber
 \end{eqnarray}

 The angle between $-\textbf{\emph{v}}_{0,\lambda_{0}}$ and $\textbf{\emph{r}}_{10,\lambda_{0}}$ is $\beta$, set the angle between $\textbf{\emph{r}}_{01,\lambda_{0}}$ and $-\textbf{\emph{v}}_{1,\lambda_{0}}$ is $\beta+\alpha.$ The analytic function of $E$ is equal to $G_{2}(\varepsilon_{1},\varepsilon_{2},p,\alpha)$ concluded from $P(a=10)=P(a=01).$

 Case 4: Let
 \begin{eqnarray}
 P(b=0|00,y=0,\lambda_{0})=p.\nonumber
 \end{eqnarray}

Set the angle between $\textbf{\emph{r}}_{00,\lambda_{0}}$ and $\textbf{\emph{v}}_{1,\lambda_{0}}$ is $\beta+\alpha.$ Denote $E$ as $G_{3}(\varepsilon_{1},\varepsilon_{2},p,\alpha)$ in this case and $\alpha\in[0,\pi-4\beta].$ By the same analysis as the case 1, we get
\begin{equation}
G_{3}(\varepsilon_{1},\varepsilon_{2},p,\alpha)=G_{1}(\varepsilon_{1},\varepsilon_{2},p,\alpha)
\end{equation}
for $\alpha\in[0,b_{1}]\cup[b_{2},\pi-4\beta].$

For $\alpha\in[b_{1}, b_{2}]$, we have
\begin{eqnarray}
G_{3}(\varepsilon_{1},\varepsilon_{2},p,\alpha)&=&\frac{1}{2}+\frac{1}{2}(\frac{1}{2}-\varepsilon_{1})^{2}(\frac{1}{2}-\varepsilon_{2})[\sigma^{2}\delta\cos\beta\nonumber\\
& &\mbox{}+\sqrt{\delta^{2}+1+2\delta\cos(2\beta+\alpha)}+\nonumber\\
& &\mbox{}\sigma^{2}\cos(\beta+\alpha)+k(\varepsilon_{1},\varepsilon_{2},p,\alpha)]\nonumber
\end{eqnarray}
where

$k(\varepsilon_{1},\varepsilon_{2},p,\alpha)=$
\begin{equation}
   \left\{
   \begin{aligned}
   & 2\sigma\sqrt{\delta^{2}+1-2\delta\cos(2\beta+\alpha)}& & if\ \alpha\in[b_{1}, a_{2})\\
   & 2\delta\sigma\cos\beta-2\sigma\cos(3\beta+\alpha)& & if\ \alpha\in [a_{2}, b_{2}].\\
   \end{aligned}
   \right.
  \end{equation}
Therefore, for $\alpha\in[b_{1}, b_{2}]$,
\begin{eqnarray}
G_{1}(\varepsilon_{1},\varepsilon_{2},p,\alpha)\geq G_{3}(\varepsilon_{1},\varepsilon_{2},p,\alpha)
\end{eqnarray}
always  is established.
We have
\begin{equation}
G(\varepsilon_{1},\varepsilon_{2},p)=\max_{\alpha\in[0,\pi-4\beta];j=1,2}\{G_{i}(\varepsilon_{1},\varepsilon_{2},p,\alpha)\}.\nonumber
\end{equation}
concluded from Eq. (C9) and Eq. (C11), where $G_{1}$ and  $G_{2}$  are shown as Eq. (C7) and Eq. (C8).

In particular, $\arcsin(\delta\sin\beta)\leq0$ and
$\pi-\arcsin(\delta\sin\beta)-\beta\geq\pi-4\beta$ can be obtained by the tool of Matlab as $\varepsilon_{1}=\varepsilon_{2}\leq 0.1358$. Then $a_{1}=0, b_{2}=\pi-4\beta$, and the set of $[0,a_{1})\cup(b_{2},\pi-4\beta]$
does not exist. The calculation process will becomes much more simple.

\nocite{*}

\bibliography{apssamp}

\begin{thebibliography}{}\label{sec:TeXbooks}
\bibitem{TeXbook} R. Colbeck, and R. Renner, Nature Physics \textbf{8}, pp. 450-453 (2012).
\bibitem{TeXbook} C. H. Bennett, and G. Brassard, in Proceedings of the IEEE International Conference on Computers, Systems and Signal Processing, Bangalore, India (IEEE, New York, 1984), pp. 175-179.
\bibitem{TeXbook} J. Bouda, M. Pivoluska, M. Plesch, and C. Wilmott1, Phys. Rev. A \textbf{86}, 062308 (2012).
\bibitem{TeXbook} R. Gallego, N. Brunner, C. Hadley, and A. Acin, Phys. Rev. Lett. \textbf{105}, 230501 (2010).
\bibitem{TeXbook} M. Paw{\l}owski and N. Brunner, Phys. Rev. A \textbf{84}, 010302(R)(2011).
\bibitem{TeXbook} R. Colbeck and A. Kent, J. Phys. A: Math. Theor. \textbf{44}, 095305 (2011).
\bibitem{TeXbook} S. Pironio \emph{et al.}, Nature (London) \textbf{464}, 1021-1024 (2010).
\bibitem{TeXbook} H-W. Li, Z-Q. Yin, Y-C. Wu, X-B. Zou, S. Wang, W. Chen, G-C. Guo, and Z-F. Han, Phys. Rev. A \textbf{84}, 034301 (2011).
\bibitem{TeXbook} H-W. Li, M. Pawlowski, Z-Q. Yin, G-C. Guo , and Z-F. Han, Phys. Rev. A \textbf{85}, 052308 (2012).
\bibitem{TeXbook} A. Ambainis, D. Leung, L. Mancinska, and M. Ozols, e-print arXiv:0810.2937.
\bibitem{TeXbook} Y-K. Wang, S-J. Qin, T-T. Song, F-Z. Guo, W. Huang, and H-J. Zuo, Phys. Rev. A \textbf{89}, 032312 (2014).
\bibitem{TeXbook} L. Masanes, e-print arXiv:0512100.
\bibitem{TeXbook} N. Nisan,  and A.Ta-Shma, J. Comput. Syst. Sci. \textbf{58}, pp. 148-173 (1999).
\bibitem{TeXbook} S. Fehr, R. Gelles, and C. Schaffner, Phys. Rev. A \textbf{87}, 012335 (2013).
\bibitem{TeXbook} J. F. Dynes, Z. L. Yuan, A. W. Sharpe, and A. J. Shields, Appl. Phys. Lett. \textbf{93}, 031109 (2008).
\bibitem{TeXbook} M. Ren, E. Wu, Y. Liang, Y. Jian, G. Wu, and H. Zeng, Phys. Rev. A \textbf{83}, 023820 (2011).





\end{thebibliography}

\end{document}